\documentclass[showpacs,prb,aps,twocolumn]{revtex4}
\usepackage{graphicx}
\usepackage[normalem]{ulem}

\usepackage{amssymb}

\begin{document}

\title{Low-Frequency Raman Modes and Electronic Excitations In Atomically Thin MoS$_{2}$ Crystals}
\author{Hualing Zeng$^{1}$}
\author{Bairen Zhu$^{1}$}
\author{Kai Liu$^{2}$}
\author{Jiahe Fan$^{2}$}
\author{Xiaodong Cui$^{1}$}
\email{xdcui@hku.hk}
\author{Q. M. Zhang$^{2}$}
\email{qmzhang@ruc.edu.cn}
\affiliation{$^{1}$Department of Physics, The University of Hong Kong, Pokfulam Road, Hong Kong, China}
\affiliation{$^{2}$Department of Physics, Renmin University of China, Beijing 100872, P. R. China}

\date{\today}

\begin{abstract}

Atomically thin MoS$_{2}$ crystals have been recognized as a quasi-2D semiconductor with remarkable physical properties. This letter reports our Raman scattering measurements on multilayer and monolayer MoS$_{2}$, especially in the low-frequency range ($<$50 cm$^{-1}$). We find two low-frequency Raman modes with contrasting thickness dependence. With increasing the number of MoS$_{2}$ layers, one shows a significant increase in frequency while the other decreases following a 1/N (N denotes the number of unit layers) trend. With the aid of first-principles calculations we assign the former as the shear mode $E_{2g}^{2}$. The latter is distinguished as the compression vibrational mode, similar to the surface vibration of other epitaxial thin films.  The opposite evolution of the two modes with thickness demonstrates novel vibrational modes in atomically thin crystal as well as a new and more precise way to characterize thickness of atomically thin MoS$_{2}$ films. In addition, we observe a broad feature around 38 cm$^{-1}$ (~5 meV) which is visible only under near-resonance excitation and pinned at the fixed energy independent of thickness. We interpret the feature as an electronic Raman scattering associated with the spin-orbit coupling induced splitting in conduction band at K points in their Brillouin zone.

\end{abstract}


\maketitle

In the hunt for ultimately thin electronic devices, atomically thin molybdenum disulfide (MoS$_{2}$) layers have been recognized as a promising material after the success in graphene. Bulk MoS$_{2}$ crystal has a structure of S-M-S covalently bonded hexagonal quasi-2D network weakly staked by weak Van der Waals forces as sketched in Figure 1a.  MoS$_{2}$ monolayer, the elementary unit to form ultrathin films by weak stacking is stable under ambient conditions and is easily fabricated either by mechanical or liquid-phase exfoliation.\cite{1,2,3} As MoS$_{2}$ thins to a few atomic layers, there emerge many exotic physical properties absent in bulk crystal.  For example MoS$_{2}$ thin film undergoes a transition from an indirect-gap semiconductor with a gap of 1.2 eV in bulk form to a direct-gap one with a band gap of 1.9 eV in monolayers,\cite{4,5} and as well as a possible structural change.\cite{Yang, Jimenez, Qin} These distinct physical properties make atomically thin MoS$_{2}$ crystal attractive for diverse applications such as electronic devices, spintronics, photovoltaic, energy storage, etc.

Raman scattering has been proven to be valuable as a nondestructive and versatile method of structural characterization and as a method for studying electronic and vibrational properties in quasi-2D materials. There are four Raman-active vibrational modes in bulk MoS$_{2}$ crystal namely E$_{1g}$ (in-plane optical vibration of S atoms), A$_{1g}$ (out-of-plane optical vibration of S atoms), $E_{2g}^{1}$ (in-plane optical vibration of Mo-S bond) and $E_{2g}^{2}$ modes (in-plane optical vibration of rigid atomic bond).\cite{6}  The E$_{1g}$ (in-plane optical vibration of S atoms) of MoS$_{2}$ thin films is negligible in the back scattering geometry which is widely used in micro-optic spectroscopic study in nanostructures, owing to the forbidden selection rule from the symmetry point view. Lee et al\cite{7} investigated the in-plane $E_{2g}^{1}$  mode and out-of-plane A$_{1g}$ mode around 400 cm$^{-1}$ in few-layer MoS$_{2}$ and found that the two modes show opposite frequency dependence on the film thickness, $E_{2g}^{1}$ softening and A$_{1g}$ stiffening with thickness presumably due to increased dielectric screening, stacking-induced changes in intralayer bonding and interlayer coupling. The frequency difference of these two characteristic $E_{2g}^{1}$ and A$_{1g}$ modes ranging from 19 cm$^{-1}$ in monolayer to 26 cm$^{-1}$ in bulk crystal are well accepted as a thickness monitor for atomically thin MoS$_{2}$ flakes. The $E_{2g}^{2}$ mode (in-plane optical vibration of rigidly bonded atoms) also known as shear mode originates from the rigid in-plane vibration against adjacent layers and seemingly exist in many layered structures. The energy of the $E_{2g}^{2}$ mode is very low,\cite{8} analogously 30-40 cm$^{-1}$ in few-layers graphene,\cite{9} and is hardly observable in many commercial apparatus due to the limited rejection against Rayleigh scattering.

\begin{figure*}
\includegraphics[angle=0,width=18cm]{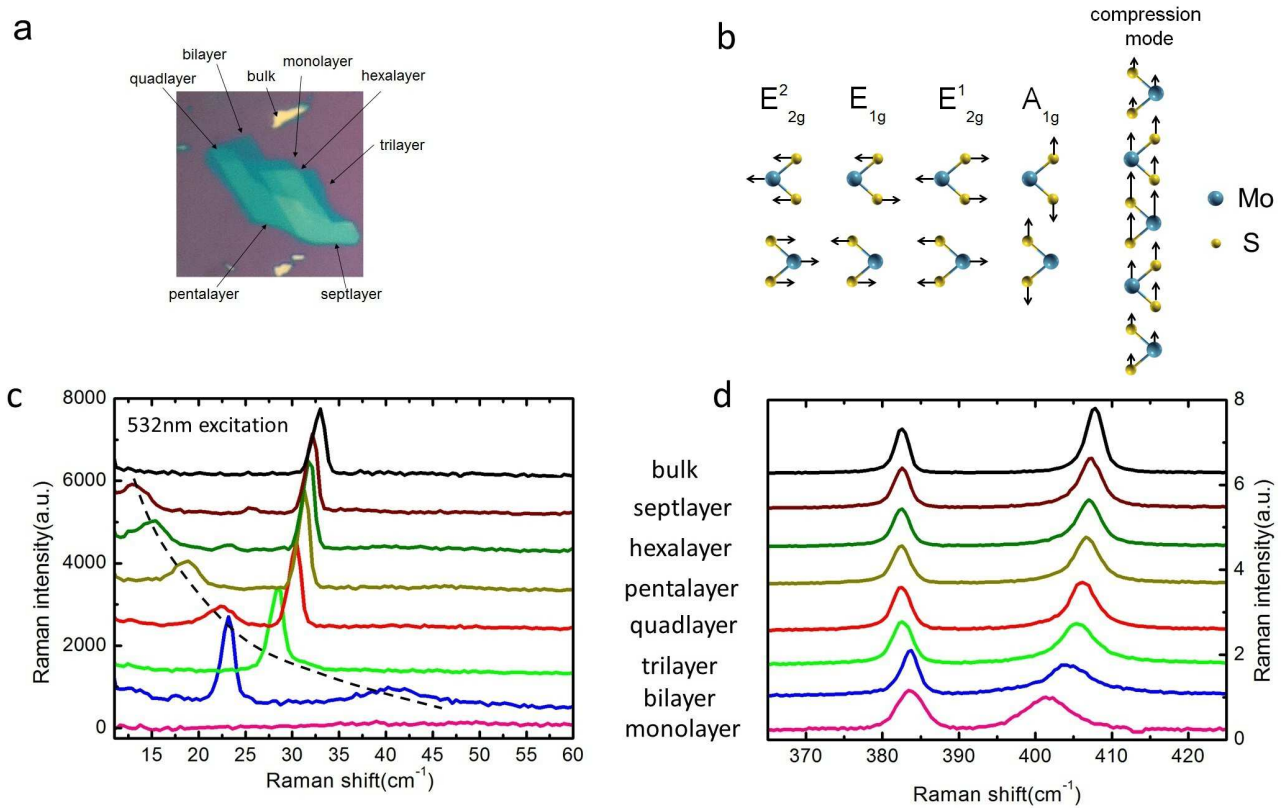}
\caption{(Color online) (a) Optical images of ultrathin MoS$_{2}$ flakes. (b) Schematics of Raman active modes. (c) Low-wavenumber Raman spectra of atomically thin MoS$_{2}$ flakes with different thickness under 532nm excitation. The dashed line denotes the evolution of the weaker low-frequency mode. (d) The corresponding E$^{1}$$_{2g}$ and A$_{1g}$ modes. Their frequency difference is an indicator of the number of layers.} \label{fig1}
\end{figure*}

Here we report our observation of Raman scattering on few-layer MoS$_{2}$ in the low frequency range. We find two low-frequency Raman modes (1$\sim$6meV) with contrasting thickness dependence. With increasing thickness of MoS$_{2}$ films, one shows a significant increase in frequency while the other decreases following a 1/N (N denotes the number of unit layers) trend. With the aid of first-principles calculations we assign the former as the shear mode $E_{2g}^{2}$ and the latter as the compression vibrational mode, similar to surface vibration of atomically epitaxial thin films. The opposite evolution of the two modes with thickness demonstrates novel vibrational modes in atomically thin crystals as well as a new way to precisely determine the number of MoS$_{2}$ atomic layers. In addition, we observe a broad feature around 5 meV, which is visible only under near-resonance excitation and locates at the fixed position at various thicknesses.

Ultrathin flakes were mechanically exfoliated from a natural MoS$_{2}$ single crystal onto silicon wafers capped with a 300nm thick SiO$_{2}$ by a method analogous to that of producing graphene. Atomically thin MoS$_{2}$ flakes were first visually screened with interference color through optical microscope. Typical optical images of MoS$_{2}$ ultrathin slabs are presented in Figure 1. The film thickness is confirmed by atomic force microscope and photo luminescence spectra.\cite{4,5} The Raman scattering is collected through a home-built confocal like microscopic set up coupling to a triple stage monochromator (Acton TriVista) equipped with a CCD camera under ambient conditions. The laser lines 532nm and 633nm are chosen to be exciting light sources with a beam spot size around 5$\sim$10 $\mu$$m^2$ at a power around 300 $\mu$W when focused at samples.

Figure 1 presents the Raman spectra from samples with various thicknesses.  The in-plane vibrational $E_{2g}^{1}$ mode and the out-of-plane vibrational A$_{1g}$ mode around 400 cm$^{-1}$ show a consistent trend as reported in Ref.7. The $E_{2g}^{1}$ mode stiffens and the A$_{1g}$ mode softens as the sample thins to multi- and monolayers. The frequency difference between the two mode shows a consistent indication of the layer thickness as other means, particularly for monolayer, bilayer, trilayer and quadlayer (N$<$4). In the low-frequency range, two modes can be clearly seen at various thicknesses except monolayer at 532nm excitation. The mode with narrow bandwidth ($\sim$2cm$^{-1}$) undergoes a blue shift with increasing thickness, from 22 cm$^{-1}$ in bilayer to 32 cm$^{-1}$ in bulk form, while the mode with broad bandwidth undergoes an opposite evolution as indicated by the dash line in Figure 1.

\begin{figure}
\includegraphics[angle=0,width=7cm]{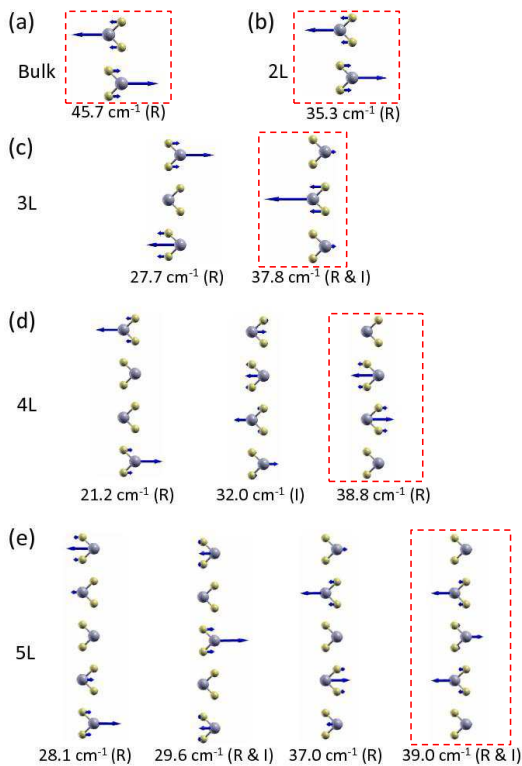}
\caption{(Color online) Calculated phonon frequencies and displacement patterns for the shear modes of (a) bulk, (b) bilayer, (c) trilayer, (d) quadlayer, and (e) pentalayer MoS$_{2}$ at the Brillouin zone center obtained from DFT calculations. The Raman-active (R) and infrared-active (I) modes are labeled. Atomic structures and displacement patterns prepared using XCRYSDEN.\cite{17a} Modes for different layers are highlighted with dashed boxes.} \label{fig2}
\end{figure}

\begin{figure}
\includegraphics[angle=0,width=7cm]{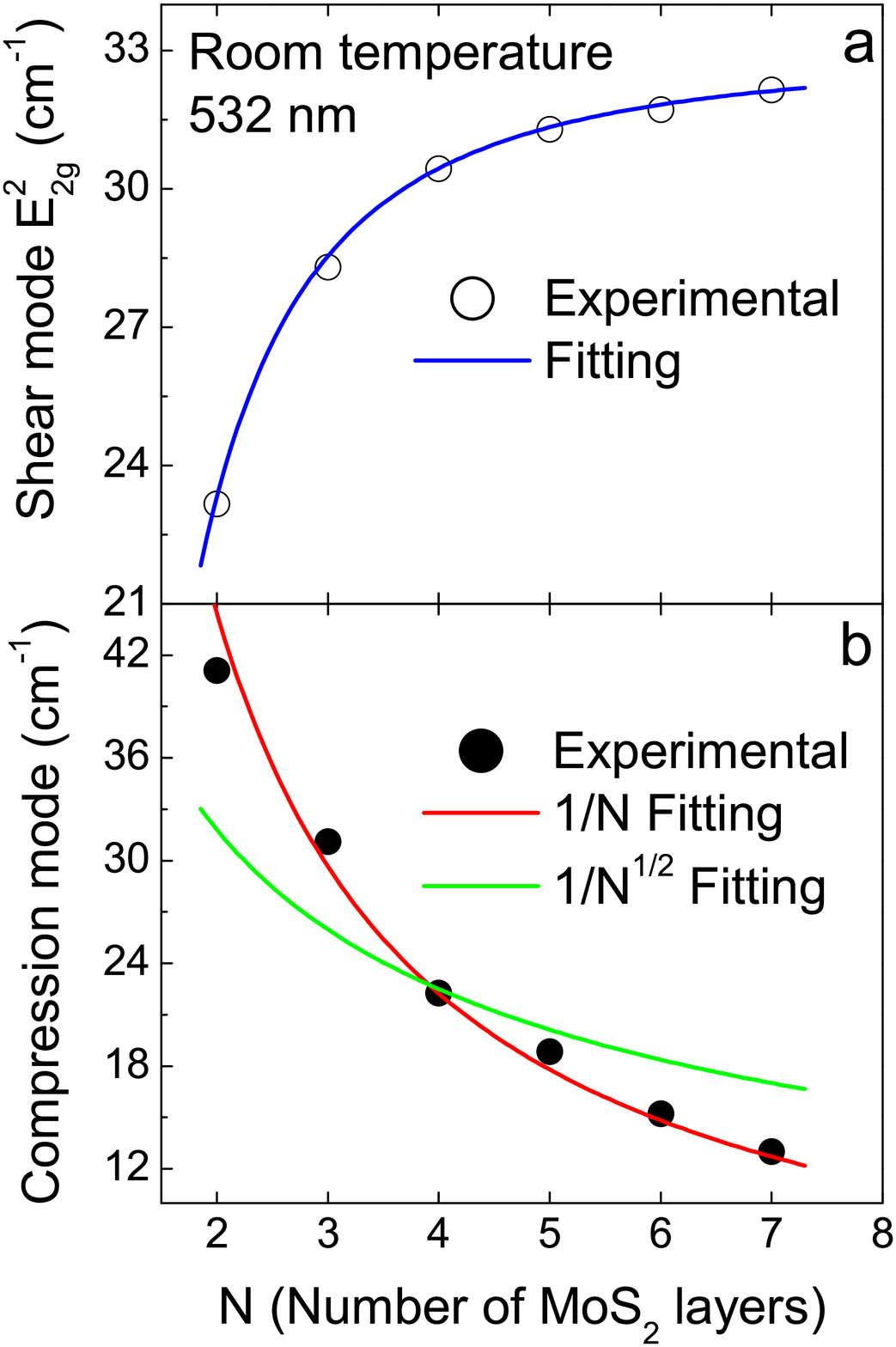}
\caption{(Color online) (a) Thickness dependence of frequencies of the $E_{2g}^{2}$ shear mode. The blue curve is the fitting with the formula $\omega=\frac{1}{\sqrt{2}\pi c}\sqrt{\frac{\alpha }{\mu }}\sqrt{1+cos(\frac{\pi }{N})}$. (b) Thickness dependence of compression mode (see below) frequencies. The red and green curves are produced with the least-square fitting.} \label{fig3}
\end{figure}

To gain insight into the observed modes, we have studied the zone center phonons of MoS$_{2}$ from first-principles density functional theory (DFT) calculations as implemented in the Vienna ab-initio simulation package.\cite{10,11,12,13} The few-layer MoS$_{2}$ were modeled as a 1$\times$1 unit cell in 2- to 5-layer slabs, which were separated by a vacuum $>$13 \r{A} in the perpendicular direction. The projector augmented wave (PAW) method\cite{14,15} and the local density approximation (LDA) for exchange-correlation potentials were adopted. The energy cutoff for the plane waves was 400 eV. The first Brillouin zone was sampled with a 12$\times$12$\times$4 k-point mesh for bulk and a 12$\times$12$\times$1 mesh for few-layers. The Gaussian broadening with a width of 0.05 eV was used for Fermi surface smearing. All ions were fully optimized until forces were less than 0.01 eV/\r{A}. The vibrational frequencies and displacement patterns of phonon modes were calculated at the equilibrium structures using the dynamical matrix method as described in our previous publication.\cite{16} It is worth noting that although the LDA method neglects the van der Waals component of the interaction between the layers, it produces reasonably well the inter-layer phonons of MoS$_{2}$.\cite{17}

The calculated frequencies of $E_{2g}^{1}$(392.7 cm$^{-1}$) and A$_{1g}$ (419.3 cm$^{-1}$) modes at the Brillouin zone center in bulk form show good agreement with our experimental values (382.9 cm$^{-1}$ and 408.2 cm$^{-1}$), as well as with previous experiments\cite{7} (382.0 cm$^{-1}$ and 407.5 cm$^{-1}$) and theoretical results (387.8 cm$^{-1}$ and 412.0 cm$^{-1}$).\cite{17} Here we concentrate on the in-plane shear modes below 50 cm$^{-1}$ which exhibit prominent thickness-dependence as shown in Figure 1. Figure 2 presents the calculated phonon frequencies and displacement patterns for the double-degenerate shear modes ($E_{2g}^{2}$ mode) in bulk and few-layer MoS$_{2}$. The calculated shear modes: 35.3 cm$^{-1}$ in bilayer(2L), 37.8 cm$^{-1}$ in trilayer(3L), 38.8 cm$^{-1}$ in quadlayer(4L) and 39.0 cm$^{-1}$ pentalayer(5L) agree qualitatively well with the corresponding observed sharp peaks around 30cm$^{-1}$ in Figure 1. The phonon frequencies of these in-plane shear modes increase from null in monolayer monotonically with the layer thickness. They show similar thickness-dependence and displacement patterns as the shear modes in multilayer graphene,\cite{9} illustrating their similar origin from the interlayer interactions. So we can safely assign the sharp peaks in Figure 1 to the shear modes ($E_{2g}^{2}$ mode as denoted in bulk) of adjacent layers oscillation driven by van der Waals interaction. Our assignments are consistent with previous reports in bulk and few-layer MoS$_{2}$.\cite{8,18}

As such a low-frequency interlayer shear mode is a genuine vibration for layered compounds, we examine the measured peak position of the shear modes with a linear-chain model\cite{9}: $\omega=\frac{1}{\sqrt{2}\pi c}\sqrt{\frac{\alpha }{\mu }}\sqrt{1+cos(\frac{\pi }{N})}$ where $\omega$, c, $\alpha$, $\mu$ and N denote the peak position of the shear mode in cm$^{-1}$, speed of light in cm/s, the interlayer force constant per unit area, unit layer mass per unit area and the number of layers, respectively. Figure 3a shows a perfect fitting with the interlayer force constant per unit area $\alpha=2.9\times10^{19} N\cdot m^{-3}$ and the single-layer mass per unit area $\mu=30.3 kg\cdot \r{A}^{-2}$. One can further estimate the shear modulus C$_{44}=\alpha \times t$ with the spacing t between adjacent MoS$_{2}$ layers is $\thicksim$ 6.2 \r{A}. This immediately gives C$_{44} \thicksim$ 17.9 GPa. The C$_{44}$ value is in good agreement with the reported experimental shear modulus in MoS$_{2}$ bulk crystal, from 15 to 19 GPa.\cite{19,20}

Now we move to the other low-frequency Raman mode. To our knowledge, this broad mode has not been reported before. As mentioned above, it cannot be assigned to any known Raman-active mode. It features a contrasting thickness-dependence against the$E_{2g}^{2}$ mode, as shown in Figure 3b. Besides, the thickness-dependence follows a perfect 1/N rather than $1/\sqrt{N}$ behavior. The signature of 1/N behavior has been observed in similar thin-film systems with a number of atomic layers grown by molecular beam epitaxy (MBE), such as Na/Cu(001).\cite{21} And it is well understood as open-ended standing wave mode vertical to atomic planes with a linear-chain force constant model, particularly including the coupling between atomic layers and substrate.\cite{22} For layered compounds, in principle there should exist a compression mode besides the shear mode discussed above, which is a rigid-layer vibration vertical to atomic planes. The longitudinal-like open standing wave will give an effective relative shift between adjacent layers if one looks at its vibration amplitude, as shown in Figure 1b. That means it actually corresponds to the compression mode. The compression mode, by combination with a longitudinal optical phonon mode, has been experimentally revealed in few-layer graphene through double-resonant intravalley scattering.\cite{23} The mode strongly depends on the coupling between atomic layers and substrate. And the Raman intensity of the mode is much smaller than that of the shear mode. This may be the reason why it is not easy to be observed in few-layer graphene and MoS$_{2}$.

The direct observation of the shear mode and the compression mode provides a new way to more precisely characterize the thickness of ultrathin MoS$_{2}$. A conventional method is to measure frequency difference between A$_{1g}$ and$E_{2g}^{1}$ mode.Considering the contrasting evolution of the two rigid-layer low-frequency modes with thickness and the large relative shift up to $100\%$, the frequency difference between the shear mode and the compression mode would be a more precise measure of layer number, particularly for slabs with more than 3 unit layers.

\begin{figure}
\includegraphics[angle=0,scale=0.3]{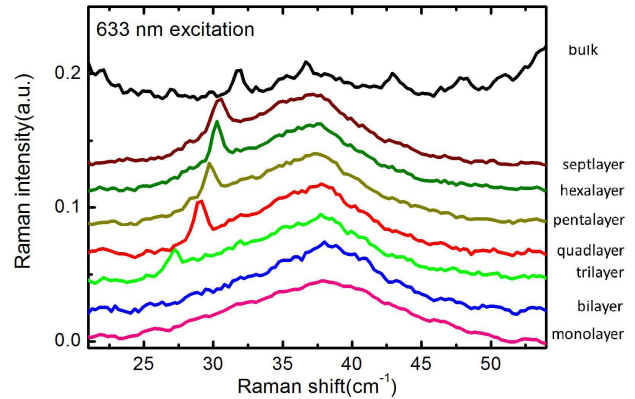}
\caption{(Color online) Low-wavenumber Raman spectra of MoS$_{2}$ thin fims at different thicknessunder near-resonant excitation (633nm). The broad feature around 38 cm$^{-1}$ is observed only in near-resonant excitation.} \label{fig4}
\end{figure}

Another new finding in our Raman measurements is the broad feature centred at 38 cm$^{-1}$ ($\thicksim$5meV) measured with both cross and parallel polarization configurations under a near-resonant excitation.Here, we present the spectrum measured in the cross polarization geometry, which shows clearer features due to the suppressed Rayleigh scattering of the exciting laser line, as shown in Figure 4. The feature does not change in either position or shape with thickness.  We can exclude first-order or higher-order phonons as its origin. Moreover, it is visible only at near-resonance excitation ($\thicksim$1.96 eV) which is very close to the direct band gap at K points in the Brillouin zone. This implies that the feature might be related to vertical electronic transition crossing direct band gap.

First the exciton mechanism could be excluded due to the energy scale difference.  The so-called A and B excitons in bulk MoS$_{2}$ are characterized with the binding energies of $\thicksim$42 and 134 meV,\cite{24} and recent optical experiments indicated that the two excitonic states exhibit little change in few-layer MoS$_{2}$ compared to bulk materials.\cite{4,5} The exciton binding energies are much larger than that of the broad feature around 5 meV. Impurity or defect states are also unlikely to contribute to this feature from the energy scale. Impurity and defects states appearing in optical measurements are of the same energy scale as the exciton binding energy.\cite{5} It should be noted that our samples and those used in the optical measurements are from the same source (SPI Supplies). The similar conditions of impurities and defects could be expected in our samples if existing. Collective electric carrier oscillation, so-called plasmon mode and other collective modes such as charge density wave (CDW) are also excluded from the fact of thickness-independence, as Coulomb screening and interlayer coupling would be strongly dependent on thickness.

The potential mechanism responsible for the broad feature may lie in the spin-orbit coupling in conduction band at K points in their Brillouin zone. Recent band structure calculations\cite{25,26} indicated that spin-orbit coupling contributes to the splitting at the top of valence band at K points ($\nu$1 and $\nu$2), which are the origins of A and B excitons. At the same time, spin-orbit coupling also induces a subtle splitting of $\thicksim$4 meV at the conduction band at K points. The subtle splitting would be reflected in a resonant-like Raman procedure crossing the direct band gap at K points. This naturally explains why it appears only at near resonant excitations. The thickness independent Raman feature manifests the thickness-independent spin-orbit coupling induced splitting in conduction band.

In conclusion, two low-frequency modes, the shear mode and the compression mode are investigated in the present Raman measurements. They exhibit completely opposite evolution with the thickness of ultrathin MoS$_{2}$. The pair of low-frequency provides a new and more precise method to characterize the thickness of atomically thin MoS$_{2}$ crystal. We also observe a broad feature at $\thicksim$5 meV which is likely related to the electronic structure. We point out that the feature may originate from the spin-orbit coupling induced splitting at the conduction band at K points in their Brillouin zone.

We thank Dr. Wang Yao and Dr. Guibin Liu for helpful discussion. This work was supported by the NSF of China (Grant Nos. 11034012 and 11004243), the 973 program (Grant Nos. 2011CBA00112 and 2012CB921701), the Fundamental Research Funds for the Central Universities, the Research Funds of Renmin University of China and research grant council of Hong Kong (GRF HKU701810P) and UGC (AoE/P-04/08 ). Computational facilities have been provided by the Physical Laboratory of High Performance Computing at Renmin University of China.

\end{document}